\documentclass[a4paper,11pt]{article}
\usepackage{graphicx}
\usepackage{epstopdf}

\usepackage[utf8]{inputenc}
\usepackage[T1]{fontenc}
\usepackage{pgfplots}
\usetikzlibrary{arrows.meta}

\usepackage{verbatim}
\usepackage{pgfplots}
\usepackage{tikz,tikz-3dplot}
\usepackage{adjustbox}

\usepackage{amsfonts,amsmath,amssymb}
\usepackage{hyperref}

\usepackage{epsfig,multicol,bbm}
 
\newcommand\fverb{\setbox\pippobox=\hbox\bgroup\verb}
\newcommand\fverbit{\egroup\item[\fbox{\unhbox\pippobox}]}

\newbox\pippobox

\begin{document}
\title{Dipole and String Solutions of the Kaluza-Klein Theory}
\author{Ahmad Farahani\thanks{Electronic address: ah.mazidabadi@mail.sbu.ac.ir}\,\, and\,\, {Nematollah Riazi\thanks{Electronic address: n\_riazi@sbu.ac.ir}}
\\
\small Department of Physics, Shahid Beheshti University, G.C., Evin, Tehran 19839, Iran}
\maketitle
\begin{abstract}
We consider a Kaluza-Klein string solution of five-dimensional spacetime. We study its physical properties as appearing in (3+1) spacetime. This metric embraces several different aspects of solutions. The two most notable examples are the dipole and the boosted dipole solutions. In general, the 4D metric is a rotating stationary solution with both electric and magnetic fields. The extension to the Kaluza-Klein dipole soliton solution is briefly discussed.
\end{abstract}
\section{Introduction}\label{sec1}
The idea of extra dimensions made its first appearance in physics by Kaluza and Klein in 1926\cite{1,2}, who attempted to unify gravitation and electromagnetism using five space-time dimensions. The theory was first suggested by Theodor Kaluza as a higher dimensional theory of gravity, and ultimately led to the eleven-dimensional supergravity theories and the ten-dimensional superstring\cite{3}. Like Einstein, Kaluza was in quest of what we call "the unified theory", that there is a theory that may explain all of the fundamental forces. He endeavored to represent the electromagnetic force in a similar way as gravity in general relativity. The existence of one extra spatial dimension with a couple of special features was suggested. The idea was this: if we want to explain one more force, maybe we need an extra dimension. Therefore, Kaluza imagined that the universe had four instead of three dimensions of space, and accordingly his theory was formulated. Since the extra dimension is not observed, if one day extra dimensions are discovered in our universe, they are probably to be compact along the lines of Kaluza-Klein theory. Several exact solutions of Kaluza-Klein equations have been discovered since the introduction of this theory\cite{2,4}.

As it is well-known, Dirac proposed a magnetic monopole with a string singularity extending from the particle's position to infinity\cite{5}. Afterwards, more magnetic monopole solutions have been developed\cite{6}. Gross and Perry\cite{7}, and Sorkin\cite{8}, simultaneously obtained a special group of solutions of the five-dimensional Kaluza-Klein theory related to a magnetic monopole. Their solutions explained a string singularity if the spatial extra dimension is compactified. Likewise, Gegenberg and Kunstatter introduced another magnetic monopole solution\cite{9}. According to\cite{10} the Kaluza-Klein monopole has an important role in M/String theory. Gross-Perry-Sorkin (GPS) magnetic monopole is a well-known solution of the Kaluza-Klein theory which is a generalization of the self-dual Euclidean Taub-NUT solution\cite{11}. This procedure can be applied to the other configurations. In a similar way, the Kaluza-Klein magnetic dipole was represented by choosing the Euclidean Kerr solution\cite{7,21}. 

In the Kaluza-Klein theory, unlike the Yang-Mills theories, there are solitons which are magnetic dipoles. The dipole configuration is made out of monopole, anti-monopole pair. The reason is that a monopole plus anti-monopole has a distinct topology from the vacuum and, moreover, can not classically annihilate. It has been claimed that the Kaluza-Klein dipole\cite{7} has a property of describing brane/anti-brane pair. Some of the dipole-like solutions with magnetic flux tubes have been constracted in \cite{22,23}. In this paper, we consider a Kaluza-Klein string solution of five-dimensional spacetime.

The structure of this paper is as follows. In section \ref{sec2}, we briefly bring up the Kaluza-Klein formalism and mention the main properties of the Kaluza-Klein dipole soliton solution. Subsequently, the boosted solution is examined. We will then present a vacuum solution of the Kaluza-Klein theory and investigate its physical properties in section \ref{sec3}. Section \ref{sec4} is devoted to summary and discussion.


\section{Kaluza-Klein Theory and the Dipole Soliton Solution}\label{sec2}
As we have already said, in this section we briefly study the Kaluza-Klein theory and then review the Kaluza-Klein dipole soliton solution. Afterwards, the boosted solution is inspected.
\subsection{Kaluza-Klein Theory: a brief review}\label{subsec2-1}
The Kaluza-Klein theory postulates the five-dimensional spacetime and the dimensional reduction of the vacuum spacetime\cite{12}, which leads to four-dimensional gravity coupled to a $U(1)$ Maxwell field and a scalar dilaton field\cite{13}. It is also assumed that the 5D energy-momentum tensor vanishes and thus
\begin{align}\label{eq1}
\hat G_{AB}=0
\end{align}
where
\begin{align}\label{eq2}
\hat G_{AB}=\hat R_{AB}-\frac{1}{2}\hat R\hat g_{AB},
\end{align}
is the 5D Einstein tensor. $\hat R_{AB}$, $\hat R$ and $\hat g_{AB}$ are the five-dimensional Ricci tensor, scalar, and metric tensor, respectively. The indices $A, B, ...$ run over $0, 1, 2, 3, 4$, and five-dimensional quantities are denoted by hats\cite{2}. The equations of motion can be derived by varying the five-dimensional Einstein-Hilbert action
\begin{align}\label{eq3}
S=-\frac{1}{16\pi\hat G}\int\hat R\sqrt{-\hat g}\:d^{4}x\:dy,
\end{align}
where $\hat G$ is the five-dimensional gravitational constant. The resulting eq.(\ref{eq2}) can be written in terms of 4D quantities. Consequently, the Kaluza-Klein field equations in four dimensions read
\begin{align}\label{eq4}
&G_{\alpha\beta}=\frac{\kappa^{2}\phi^{2}}{2}T^{EM}_{\alpha\beta}-\frac{1}{\phi}[\nabla_{\alpha}(\partial_{\beta}\phi)-g_{\alpha\beta}\Box\phi],\\
&\nabla^{\alpha}F_{\alpha\beta}=-3\frac{\partial^{\alpha}\phi}{\phi}F_{\alpha\beta},\\
&\Box\phi=\frac{\kappa^{2}\phi^{3}}{4}F_{\alpha\beta}F^{\alpha\beta},
\end{align}
where $T^{EM}_{\alpha\beta}=\frac{1}{4}g_{\alpha\beta}F_{\mu\nu}F^{\mu\nu}-F_{\alpha}\!^{\mu}F_{\beta\mu}$ is the electromagnetic energy-momentum tensor, and $F_{\alpha\beta}=\partial_{\alpha}A_{\beta}-\partial_{\beta}A_{\alpha}$ is the field strength. $\phi$ and $\kappa$ are the scalar field and coupling constant for the electromagnetic potential $A_{\alpha}$, respectively\cite{14,15}. (Throughout this paper, Greek indices $\alpha, \beta, ...$ run over $0, 1, 2, 3$).


\subsection{Dipole Soliton Solution}\label{subsec2-2}
The Kaluza-Klein dipoles are described by the Kerr-Schwarzschild metrics\cite{16} which can be obtained from the Kerr-Taub-Bolt solutions\cite{17} which contain both elementry monopoles and elementry dipoles. The Kaluza-Klein dipoles are regular solutions for $3+1$ spatial dimension in which the fourth dimension is periodic. The line element is represented by the following metric\cite{18}
\begin{align}\label{eq8}
ds^{2}=&-dt^{2}+\frac{1}{r^{2}-a^{2}\cos^{2}\theta}\big[\Delta(dy+a\sin^{2}\theta d\psi)^{2}+\sin^{2}\theta((r^{2}-a^{2})d\psi-ady)^{2}\big]\nonumber\\
&+(r^{2}-a^{2}\cos^{2}\theta)\big[\frac{dr^{2}}{\Delta}+d\theta^{2}\big],
\end{align}
where
\begin{align}\label{eq9}
&\Delta=r^{2}-2mr-a^{2},\\
&\psi=\phi+\Omega y,
\end{align}
and $y$ is identified with period $2\pi/\kappa$. Here
\begin{align}\label{eq9-1}
\Omega=\kappa\frac{a}{\sqrt{m^{2}+a^{2}}},\ \ \ \ \ \kappa=\frac{\sqrt{m^{2}+a^{2}}}{2m[m+\sqrt{m^{2}+a^{2}}]}.
\end{align}
This solution describes a magnetic dipole. When $r\to\infty$, the vector potential is given by
\begin{align}\label{eq9-2}
A_{\phi}\sim\frac{-2ma\sin^{2}\theta}{r}.
\end{align}
The field of the magnetic dipole is pointing along the $z$ axis. This dipole is not produced by rotating currents and it has zero angular momentum. Moreover, the dipole mass is determind by the value of $a$. There is only one fixed point of the Killing vector $\frac{\partial}{\partial y}$, which is at the source of the magnetic dipole, $r=m+\sqrt{m^{2}+a^{2}}$\cite{7}.

\subsubsection{The Boosted Kaluza-Klein Dipole}\label{subsubsec2-3}
From the five-dimensional standpoint, we implement a boost to the Kaluza-Klein dipole. The proposed boost is along the extra spatial dimension $y$. We determine the boosted coordinates as ($t$, $r$, $\theta$, $\phi$, $y$), and consider metric (\ref{eq8}) with coordinate renamed as ($t^{\prime}$, $r^{\prime}$, $\theta^{\prime}$, $\phi^{\prime}$, $y^{\prime}$). Accordingly, we apply the following transformations
\begin{align}\label{eq10}
&t^{\prime}=t\cosh\alpha-y\sinh\alpha,\\
&y^{\prime}=y\cosh\alpha-t\sinh\alpha,
\end{align}
where $\alpha$ is the boosted parameter. So, the transformed metric is given by
\begin{align}\label{eq11}
ds^{2}=&-\Big[\cosh^{2}\alpha-\frac{1}{r^{2}-a^{2}\cos^{2}\theta}\big((r^{2}-2mr-a^{2})(1+a\Omega\sin^{2}\theta)^{2}\sinh^{2}\alpha\nonumber\\
&+(r^{2}\Omega-a^{2}\Omega-a)^{2}\sin^{2}\theta\sinh^{2}\alpha\big)\Big]dt^{2}+\big(\frac{r^{2}-a^{2}\cos^{2}\theta}{r^{2}-2mr-a^{2}}\big)dr^{2}+\big(r^{2}-a^{2}\cos^{2}\theta\big)d\theta^{2}\nonumber\\
&+\Big[\frac{(r^{2}-2mr-a^{2})}{r^{2}-a^{2}\cos^{2}\theta}a^{2}\sin^{4}\theta+\frac{(r^{2}-a^{2})^{2}\sin^{2}\theta}{r^{2}-a^{2}\cos^{2}\theta}\Big]d\phi^{2}+\Big[-\sinh^{2}\alpha\nonumber\\
&+\frac{\cosh^{2}\alpha}{r^{2}-a^{2}\cos^{2}\theta}\big((r^{2}-2mr-a^{2})(1+a\Omega\sin^{2}\theta)^{2}+(r^{2}\Omega-a^{2}\Omega-a)^{2}\sin^{2}\theta\big)\Big]dy^{2}\nonumber\\
&+\Big[\frac{1}{r^{2}-a^{2}\cos^{2}\theta}\big((r^{2}-2mr-a^{2})(-2\cosh\alpha\sinh\alpha)(1+a\Omega\sin^{2}\theta)^{2}\nonumber\\
&-2(r^{2}\Omega-a^{2}\Omega-a)^{2}\cosh\alpha\sinh\alpha\sin^{2}\theta\big)+2\cosh\alpha\sinh\alpha\Big]dtdy\nonumber\\
&+\Big[\frac{1}{r^{2}-a^{2}\cos^{2}\theta}\big(2a^{2}\sin^{4}\theta(r^{2}-2mr-a^{2})(1+a\Omega\sin^{2}\theta)\nonumber\\
&-2(r^{2}-a^{2})(r^{2}\Omega-a^{2}\Omega-a)\sin^{2}\theta\cosh\alpha\big)\Big]d\phi dy+\Big[\frac{1}{r^{2}-a^{2}\cos^{2}\theta}\big(-2a^{2}\sin^{4}\theta\sinh\alpha(r^{2}\nonumber\\
&-2mr-a^{2})(1+a\Omega\sin^{2}\theta)-(r^{2}\Omega-a^{2}\Omega-a)\sin^{2}\theta\sinh\alpha\big)\Big]dtd\phi
\end{align}
Because of the $dtd\phi$ term the metric becomes a stationary rather than static solution. The transformed scalar and gauge fields resulting from the above metric are given by
\begin{align}\label{eq12}
\phi^{2}=-\sinh^{2}\alpha+\frac{\cosh^{2}\alpha}{r^{2}-a^{2}\cos^{2}\theta}\big[(r^{2}-2mr-a^{2})(1+a\Omega\sin^{2}\theta)^{2}+(r^{2}\Omega-a^{2}\Omega-a)^{2}\sin^{2}\theta\big],
\end{align}
\begin{align}\label{eq13}
A_{t}=&\frac{1}{\kappa}\big[\frac{(r^{2}-2mr-a^{2})(-2\cosh\alpha\sinh\alpha)(1+a\Omega\sin^{2}\theta)^{2}}{-\sinh^{2}\alpha(r^{2}-a^{2}\cos^{2}\theta)+\cosh^{2}\alpha\big((r^{2}-2mr-a^{2})(1+a\Omega\sin^{2}\theta)^{2}+(r^{2}\Omega-a^{2}\Omega-a)^{2}\sin^{2}\theta\big)}\nonumber\\
&+\frac{-2(r^{2}\Omega-a^{2}\Omega-a)^{2}\cosh\alpha\sinh\alpha\sin^{2}\theta+2\cosh\alpha\sinh\alpha}{-\sinh^{2}\alpha(r^{2}-a^{2}\cos^{2}\theta)+\cosh^{2}\alpha\big((r^{2}-2mr-a^{2})(1+a\Omega\sin^{2}\theta)^{2}+(r^{2}\Omega-a^{2}\Omega-a)^{2}\sin^{2}\theta\big)}\big],
\end{align}
\begin{align}\label{eq14}
A_{\phi}=&\frac{1}{\kappa(r^{2}-a^{2}\cos^{2}\theta)}\times\nonumber\\
&\frac{2a^{2}\sin^{4}\theta(r^{2}-2mr-a^{2})(1+a\Omega\sin^{2}\theta)-2(r^{2}-a^{2})(r^{2}\Omega-a^{2}\Omega-a)\sin^{2}\theta\cosh\alpha}{-\sinh^{2}\alpha(r^{2}-a^{2}\cos^{2}\theta)+\cosh^{2}\alpha\big((r^{2}-2mr-a^{2})(1+a\Omega\sin^{2}\theta)^{2}+(r^{2}\Omega-a^{2}\Omega-a)^{2}\sin^{2}\theta\big)},
\end{align}
which means the boosted Kaluza-Klein dipole also has an electric field. It can be reduced to the early metric (\ref{eq8}) by setting $\alpha=0$.


\section{The Solution}\label{sec3}
Here, we introduce another metric which is a vacuum five-dimensional solution, having some properties in common with the string solution. The proposed stationary metric is given by
\begin{align}\label{eq19}
ds^{2}=&-\left(1-C^{2}r^{2}\sin^{2}\theta\right)dt^{2}+\left(1+I^{2}(r)r^{2}\sin^{2}\theta\right)dr^{2}+r^{2}d\theta^{2}+r^{2}\sin^{2}\theta d\phi^{2}\nonumber\\
&+\left(1+A^{2}r^{2}\sin^{2}\theta\right)dy^{2}+2CI(r)r^{2}\sin^{2}\theta dtdr+2Cr^{2}\sin^{2}\theta dtd\phi+2CAr^{2}\sin^{2}\theta dtdy\nonumber\\
&+2I(r)r^{2}\sin^{2}\theta drd\phi+2AI(r)r^{2}\sin^{2}\theta drdy+2Ar^{2}\sin^{2}\theta d\phi dy,
\end{align}
where the extra spatial coordinate is represented by $y$, and $A$ and $C$ are constants. $I(r)$ is an arbitrary function of $r$. The coordinates are given by $r$, $\theta$, $\phi$ with usual ranges $r\ge0$, $0\le\theta\le\pi$, $0\le\phi\le2\pi$ and $0\le y\le2\pi$. The metric has the signature $(-++++)$. We have constructed the metric (\ref{eq19}) simply by boosting and transforming the $4+1$ flat solution in such a way that it satisfies the vacuum Einstein equations in five dimensions while keeping all metric parameters, although some are obviously superficial.

The scalar field $\phi$, and the gauge field $A_{\mu}$ deduced from the metric (\ref{eq19}) are
\begin{align}\label{eq20}
\phi^{2}=1+A^{2}r^{2}\sin^{2}\theta,
\end{align}
and
\begin{align}\label{eq21}
A_{t}=\frac{1}{\kappa}\frac{CAr^{2}\sin^{2}\theta}{\left(1+A^{2}r^{2}\sin^{2}\theta\right)},
\end{align}
\begin{align}\label{eq22}
A_{r}=\frac{1}{\kappa}\frac{AI(r)r^{2}\sin^{2}\theta}{\left(1+A^{2}r^{2}\sin^{2}\theta\right)},
\end{align}
\begin{align}\label{eq23}
A_{\phi}=\frac{1}{\kappa}\frac{Ar^{2}\sin^{2}\theta}{\left(1+A^{2}r^{2}\sin^{2}\theta\right)},
\end{align}
respectively.

The size of the Kaluza-Klein circle is as follows 
\begin{align}\label{eq23-1}
C_{KK}=\int_{0}^{2\pi}\sqrt{1+A^{2}r^{2}\sin^{2}\theta}dy=2\pi\sqrt{1+A^{2}r^{2}\sin^{2}\theta},
\end{align}
which it depends on the values of $A$ and $r$ as well as $\theta$. Thus, the size of the  Kaluza-Klein circle becomes non-compactified at large $r$ when $A\neq0$ , and hence the interpretation of the solution as a  Kaluza-Klein breaks down at large distance. Therefore, in this paper we concentrate at small $r$ to satisfy the  Kaluza-Klein reduction.

Moreover, the components of the electromagnetic fields are given by
\begin{align}\label{eq24}
F_{tr}=-\frac{1}{\kappa}\frac{2ACr\sin^{2}\theta}{(1+A^{2}r^{2}\sin^{2}\theta)^{2}}=E_{r},
\end{align}
\begin{align}\label{eq24-1}
F_{t\theta}=-\frac{1}{\kappa}\frac{2ACr^{2}\sin\theta\cos\theta}{(1+A^{2}r^{2}\sin^{2}\theta)^{2}}=rE_{\theta},
\end{align}
\begin{align}\label{eq25}
F_{\theta\phi}=\frac{1}{\kappa}\frac{2Ar^{2}\sin\theta\cos\theta}{(1+A^{2}r^{2}\sin^{2}\theta)^{2}}=-r^{2}\sin\theta B_{r},
\end{align}
\begin{align}\label{eq26}
F_{r\phi}=\frac{1}{\kappa}\frac{2Ar\sin^{2}\theta}{(1+A^{2}r^{2}\sin^{2}\theta)^{2}}=r\sin\theta B_{\theta},
\end{align}
\begin{align}\label{eq27}
F_{r\theta}=-\frac{1}{\kappa}\frac{2AI(r)r^{2}\sin\theta\cos\theta}{(1+A^{2}r^{2}\sin^{2}\theta)^{2}}=-rB_{\phi}.
\end{align}
The three-dimensional electric and magnetic field lines are shown in Figs. (\ref{Fig.1}) and (\ref{Fig.2}), respectively. By converting the magnetic fields from spherical coordinates to a cartesian one and setting $I(r)=0$, we will have a magnetic field along the $z$ axis. An assosiation of the magnetic field components $B_{r}$, $B_{\theta}$ and $B_{z}$ are given by
\begin{align}\label{eq27-1}
B_{r}^{2}+B_{\theta}^{2}=B^{2}_{z}|_{I(r)=0}=\frac{4A^{2}}{\kappa^{2}\phi^{2}}.
\end{align}
We can easily show that
\begin{align}\label{eq27-2}
\overrightarrow{\nabla}.\overrightarrow{B}=0\ .
\end{align}
\begin{figure}
\begin{adjustbox}{center}
\begin{tikzpicture}
\begin{axis}[xlabel=$x$, ylabel=$y$, zlabel=$z$, view={60}{120},
domain=-1:1,
xmax=1,
ymax=1,
]
\addplot3[cyan,/pgfplots/quiver,
quiver/u=y,
quiver/v=z,
quiver/w=x,
quiver/scale arrows=0.25,
-stealth,samples=10] ({-2*x/(1+x^2+y^2)^2},{-2*y/(1+x^2+y^2)^2},{0});
\end{axis}
\end{tikzpicture}
\end{adjustbox}
\caption{The three-dimensional electric field lines for $A=C=1$}\label{Fig.1}
\end{figure}
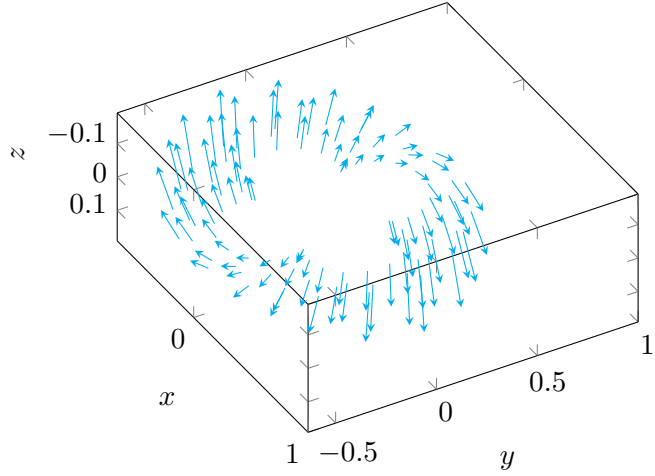

\begin{figure}
\begin{adjustbox}{center}
\begin{tikzpicture}
\begin{axis}[xlabel=$x$, ylabel=$y$, zlabel=$z$, view={135}{175},
domain=-0.3:0.3,
xmax=0.3,
ymax=0.3,
]
\addplot3[cyan,/pgfplots/quiver,
quiver/u=y,
quiver/v=z,
quiver/w=x,
quiver/scale arrows=0.015,
-stealth,samples=10] ({-4*y*(1/1-(1+x^2+y^2)^1/2)/(1+4*x^2+4*y^2)^2*(1+x^2+y^2)^1/2},{4*x*(1/1-(1+x^2+y^2)^1/2)/(1+4*x^2+4*y^2)^2*(1+x^2+y^2)^1/2},{-4/(1+4*x^2+4*y^2)^2});
\end{axis}
\end{tikzpicture}
\end{adjustbox}
\caption{The three-dimensional magnetic field lines for $A=1$ and $I(r)=1/(1-r)$. As we pointed out, $I(r)$ is an arbitrary function of $r$.}\label{Fig.2}
\end{figure}
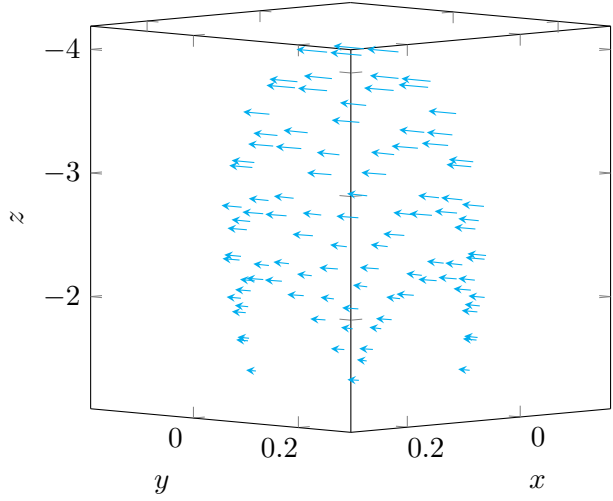

The four-dimensional spacetime is described by the following metric which is obtained by performing a Kaluza-Klein reduction
\begin{align}\label{eq28}
ds^{2}_{4D}=&-\Big(1-\big(\frac{C^{2}}{A^{2}}\big)\frac{1}{1+\frac{1}{A^{2}r^{2}\sin^{2}\theta}}\Big)dt^{2}+\Big(1+\big(\frac{I^{2}(r)}{A^{2}}\big)\frac{1}{1+\frac{1}{A^{2}r^{2}\sin^{2}\theta}}\Big)dr^{2}+r^{2}d\theta^{2}\nonumber\\
&+\big(\frac{1}{A^{2}}\big)\frac{1}{1+\frac{1}{A^{2}r^{2}\sin^{2}\theta}}d\phi^{2}+\big(\frac{CI(r)}{A^{2}}\big)\frac{1}{1+\frac{1}{A^{2}r^{2}\sin^{2}\theta}}dtdr+\big(\frac{C}{A^{2}}\big)\frac{1}{1+\frac{1}{A^{2}r^{2}\sin^{2}\theta}}dtd\phi\nonumber\\
&+\big(\frac{I(r)}{A^{2}}\big)\frac{1}{1+\frac{1}{A^{2}r^{2}\sin^{2}\theta}}drd\phi,
\end{align}
The inverse metric tensor is given by
\begin{align*}
g^{\alpha\beta}=
\begin{pmatrix}
-1 & 0 & 0 & C \\
0 & 1 & 0 & -I(r) \\
0 & 0 & 1/r^{2} & 0 \\
C & -I(r) & 0 & H(r,\theta)
\end{pmatrix}
,
\end{align*}
where
\begin{equation}\label{eq29-1}
H(r,\theta)= \frac{1}{r^{2}\sin^{2}\theta}+A^{2}+I^{2}(r)-C^{2}.\nonumber
\end{equation}
Accordingly, the Ricci scalar $R$, and the nontrivial quadratic curvature invariant $R^{\alpha\beta\mu\nu}R_{\alpha\beta\mu\nu}$, and the curvature singularities are determined
\begin{align}\label{eq29-1}
r_{1}=&\frac{1}{A}\sqrt{\frac{2(\cos^{2}\theta-1)}{\cos^{4}\theta-2\cos^{2}\theta+1}}\ ,\\
r_{2}=&-\frac{1}{A}\sqrt{\frac{2(\cos^{2}\theta-1)}{\cos^{4}\theta-2\cos^{2}\theta+1}}\ ,\\
r_{3}=&\left(-\frac{\cos^{12}\theta-6\cos^{10}\theta+6\cos^{8}\theta+16\cos^{6}\theta-39\cos^{4}\theta+30\cos^{2}\theta-8}{\left(\cos^{6}\theta-3\cos^{4}\theta+3\cos^{2}\theta-1\right)^{3}}\right)^{1/3}\nonumber\\
&+\frac{1}{3}\left(\frac{3\cos^{4}\theta-6\cos^{2}\theta+3}{\cos^{6}\theta-3\cos^{4}\theta+3\cos^{2}\theta-1}\right)\ .
\end{align}
Note that $r_{2}(\theta)$ is irrelevant since it is negative, thus not physical. In general, $r_{1}$ is imaginary in $\theta\in(0,\pi)$, and if $\theta=0,\pi$ then $r_{1}\to+\infty$ which is beyond our near-origin investigation. It is easy to understand that $r_{3}$ is a negative function of coordinate $\theta$. However, if $\theta=0, \pi$, then $r_{3}\to+\infty$. Therefore, a string singularity is present.

The four-dimensional metric has at least two Killing vector $\partial_{t}$ and $\partial_{\phi}$. In order to obtain the Killing horizon, we can use the condition $\xi^{2}=0$, where $\xi$ is an otherside timelike Killing vector, therefore we have
\begin{align}\label{eq30}
g_{\mu\nu}\xi^{\mu}\xi^{\nu}=-1+\frac{C^{2}r^{2}\sin^{2}\theta}{1+A^{2}r^{2}\sin^{2}\theta}=0,
\end{align}
which gives
\begin{align}\label{eq30}
r=\frac{1}{\sin\theta}\frac{1}{\sqrt{C^{2}-A^{2}}},
\end{align}
that is similar to the infinite red-shift surface $g_{tt}=0$. Furthermore, the event horizon can be obtained by $g^{rr}=0$. However, for metric (\ref{eq28}) $g^{rr}=1$, which means there is no event horizon.

According to the radial electric and magnetic fields, we can calculate the net electric and magnetic fluxes through any two-dimensional surface. The electric flux can be calculated through\cite{19}
\begin{align}\label{eq34}
Q_{E}=-\int_{\partial\Sigma}{d^{n-2}z\sqrt{|\gamma^{\partial\Sigma}|}n_{\mu}\sigma_{\nu}F^{\mu\nu}}\ ,
\end{align}
where $\Sigma$ is a hypersurface of constant $t$ and $r$ and, $|\gamma^{\partial\Sigma}|$ is the determinant of the induced metric on $\partial\Sigma$, $n_{\mu}$ and $\sigma_{\nu}$ are the unit normal vectors are given by
\begin{align}\label{eq34-1}
n^{\mu}=(1,0,0,0)\ \ ,\ \ \sigma^{\mu}=(0,1,0,0),
\end{align}
therefore
\begin{align}\label{eq34-2}
Q_{E}=-\lim_{r\to\infty}\int_{s^{2}}r^{2}\sin(\theta)d\theta d\phi n^{t}\sigma^{r}F_{tr}.
\end{align}
After some calculation the electric flux is shown to be zero\footnotemark
\begin{align}\label{eq35}
Q_{E}=0.
\end{align}
The magnetic flux for metric (\ref{eq28}) is as follows
\begin{align}\label{eq36}
\Phi_{B}=-\int_{\partial\Sigma}{d^{n-2}z\sqrt{|\gamma^{\partial\Sigma}|}n_{\mu}\sigma_{\nu}\star F^{\mu\nu}}\ ,
\end{align}
where
\begin{align}\label{eq36-1}
\star F^{\mu\nu}=\frac{1}{2}\epsilon^{\mu\nu\rho\sigma}F_{\rho\sigma},
\end{align}
thus
\begin{align}\label{eq37}
\Phi_{B}=-\frac{2\pi A}{\kappa}\lim_{r\to\infty}\int_{0}^{\pi}\frac{r^{4}\sin^{2}\theta\cos\theta}{(1+A^{2}r^{2}\sin^{2}\theta)^{2}}d\theta,
\end{align}
so that the net magnetic flux vanishes, too\footnotemark[\value{footnote}]\footnotetext{Note that the value of the integral becomes zero before we apply $r\to\infty$.} 
\begin{align}\label{eq39}
\Phi_{B}=0.
\end{align}
We conclude that the solution is not a magnetic monopole.

Because of the existence of two Killing vectors $\xi^{\mu}=\delta^{\mu}_{t}$ and $\xi^{\mu}=\delta^{\mu}_{\phi}$ we can get the mass $M$ and the angular momentum $J$, which correspond to the time translation and the axial symmetry, respectively. To calculate the conserved quantities we use the following integral\cite{20}
\begin{align}\label{eq40}
I=\frac{1}{8\pi G}\int_{s}\nabla^{n}\xi^{m}d^{2}\Sigma_{mn}\ ,
\end{align}
where $d^{2}\Sigma_{mn}$ is a two-dimensional surface. By using
\begin{align}\label{eq40-1}
d^{2}\Sigma_{mn}=\epsilon_{mn\theta\phi}r^{2}\sin\theta d\theta d\phi,
\end{align}
we can easily show that the relevant integration measure for the time translation is as follows
\begin{align}\label{eq40-2}
\nabla^{n}\xi^{m}d^{2}\Sigma_{mn}=2\nabla ^{r}\xi^{t}r^{2}\sin\theta d\theta d\phi.
\end{align}
Substituting into the integral we obtain
\begin{align}\label{eq41}
M=\frac{1}{8\pi G}\int 2\nabla ^{r}\xi^{t}r^{2}\sin\theta d\theta d\phi=0.
\end{align}
In the case of axial symmetry the integral will give $J=0$.


\subsection{Investigation of Special Cases}\label{subsec3}
We now examine the solution for special cases of constant parameters.

Take, for instance, $C=0$ and assuming a well-behaved $I(r)$, we can look for $I(r_{0})=0$, in which $r_{0}$ is a root of the function $I(r)$. In this case, the metric (\ref{eq19}) becomes static. The electric field becomes zero, the magnetic field is given by Eqs. (\ref{eq25}) and (\ref{eq26}), and the scalar dilaton field is (\ref{eq20}). If $\theta=0,\pi$, then $B_{r}\neq0$ and $B_{\theta}=0$. If $\theta=\frac{\pi}{2}$, then $B_{r}=0$ and $B_{\theta}\neq0$. Consequently, the infinite red-shift surface besomes imaginary and the carvature singularity is a magnetic string singularity. The Kaluza-Klein dipole solution (\ref{eq8}) \big[or eq.(30) in \cite{7}\big] behaves this case.

In the $C=0$ case, the electric field is zero and there are three components of the magnetic field in the spherical coordinates. The scalar dilaton field is given by eq. (\ref{eq20}). Once again, the metric (\ref{eq19}) becomes static. The event horizon and the infinite red-shift surface do not exist. Taking $\theta=0,\frac{\pi}{2}, \pi$, $B_{\phi}$ becomes zero and the other magnetic field components behave like the previous situation.

By applying $A=0$, we see that there are no electric and magnetic fields and the scalar diaton field would be constant. The event horizon wouldn't exist, therefore the four-dimensional spacetime singularity is naked. The infinite red-shift surface is as follows
\begin{align}\label{eq43}
r=\frac{1}{C\sin^{2}\theta}.
\end{align}
We conclude that, there is no magnetic string and the carvature singularity is still a string singularity along $\theta=0,\pi$.

By putting $I(r)=0$, $B_{\phi}$ becomes zero. Other components of the magnetic and electric fields exist. In this case, the metric (\ref{eq19}) remains stationary and behaves like the boosted Kaluza-Klein dipole (\ref{eq11}).

\section{Conclusion}\label{sec4}
We considered a Kaluza-Klein string solution which included both dipole and boosted dipole soliton solutions for special cases of parameters which appear in the solution. This solution was studied for small $r$ when $A\neq0$, where the Kaluza-Klein reduction does not break down. It is also not irrelevant to investigate this solution at large $r$ when $A\neq0$ which leads to the non-compactified Kaluza-Klein approach. The substantial contribution of this paper was the introduction of this solution and investigating the physical properties of the represented solution. The gravitational mass was calculated and shown to vanish. We computed the magnetic charge and demonstrated that the net magnetic flux of the solution would be zero, which means that there is no extended monopole source. The three-dimensional electric and magnetic fields lines were drawn. In general, it was pointed out that the carvature singularity is not covered by a horizon. It was also shown that the infinite red-shift surface is associated with the $A$ and $C$ parameters. As a special case by applying $|C|<|A|$, we figured out that there is no infinite red-shift surface.


\end{document}